\documentclass[useAMS,usenatbib]{mn2e}
\usepackage{aas_macros}
\usepackage[a4paper,centering, totalwidth=520pt, totalheight=700pt]{geometry}
\usepackage{graphicx}
\usepackage{amssymb}
\usepackage{amsmath}
\usepackage{enumerate}

\usepackage{color}

\newcommand{\del}{\ensuremath{\delta}}

\newcommand{\gam}{\ensuremath{\gamma}}

\newcommand{\sig}{\ensuremath{\sigma}}

\def\Mpc{\, h^{-1}{\rm Mpc}}

\newcommand{\avg}[1]{\ensuremath{\left\langle \,#1\, \right\rangle}}

\newcommand{\der}{\ensuremath{{\rm d}}}
\newcommand{\dir}{\ensuremath{\delta_{\rm D}}}

\newcommand{\erf}[1]{\ensuremath{{\rm erf}\left(#1\right)}}

\newcommand{\eqn}[1]{equation~\eqref{#1}}

\newcommand{\ph}[1]{\phantom{#1}}

\newcommand{\be}{\begin{equation}}
\newcommand{\ee}{\end{equation}}

\definecolor{ForestGreen}{rgb}{0.3,0.7,0.3}

\newcommand{\zobov}{{\scshape zobov}}

\title[Void hierarchy in dark matter]
      {Testing spherical evolution for modelling void abundances}
\author[I. Achitouv et al.]
{Ixandra Achitouv$^{1,2}$\thanks{E-mail:achitouv@usm.lmu.de },
  Mark Neyrinck$^3$\thanks{E-mail:neyrinck@pha.jhu.edu}
  \& Aseem Paranjape$^4$\thanks{E-mail: aseemp@phys.ethz.ch}\\
 $^1$ Ludwig-Maximilians-Universitat M\"unchen, Universit\"ats-Sternwarte M\"unchen, Scheinerstr. 1, D-81679 M\"unchen, Germany\\
$^2$Excellence Cluster Universe, Boltzmannstr. 2, D-85748 Garching bei M\"unchen, Germany\\
 $^3$ Department of Physics and Astronomy, The Johns Hopkins University, Baltimore, Maryland 21218, USA\\
 $^4$ Institute for Astronomy, Department of Physics, ETH Z\"urich, Wolfgang-Pauli-Strasse 27, CH-8093 Z\"urich, Switzerland
}
\date{}

\begin{document}
\pagerange{\pageref{firstpage}--\pageref{lastpage}}

\maketitle 

\label{firstpage}

\begin{abstract}
We compare analytical predictions of void volume functions to those measured from $N$-body simulations, detecting voids with the \zobov\ void finder. We push to very small, nonlinear voids, below few $h^{-1}$.Mpc radius, by considering the unsampled DM density field. We also study the case where voids are identified using halos. We develop analytical formula for the void abundance of both the excursion set approach and the peaks formalism. These formula are valid for random walks smoothed with a top-hat filter in real space, with a large class of realistic barrier models. We test the extent to which the spherical evolution approximation, which forms the basis of the analytical predictions, models the highly aspherical voids that occur in the cosmic web, and are found by a watershed-based algorithm such as \zobov. 
We show that the volume function returned by \zobov\ is quite sensitive to the choice of treatment of sub-voids, a fact that has not been appreciated previously. 
For reasonable choices of sub-void exclusion, we find that the Lagrangian density $\delta_v$ of the \zobov\ voids -- which is predicted to be a constant $\delta_v\approx-2.7$ in the spherical evolution model -- is different from the predicted value, showing substantial scatter and scale dependence. This result applies to voids identified at $z=0$ with effective radius between 1 and 10$Mpc.h^{-1}$.
Our analytical approximations are flexible enough to give a good description of the resulting volume function; however, this happens for choices of parameter values that are different from those suggested by the spherical evolution assumption.
We conclude that analytical models for voids must move away from the spherical approximation in order to be applied successfully to observations, and we discuss some possible ways forward.

\end{abstract}


\section{Introduction}

A visually striking aspect of all galaxy surveys to date is the presence of large, nearly empty regions known as \emph{voids} \citep{koss81,kf91,hv02,hv04,Croton04,p+06,Pan12,Sutter12}. There has been considerable interest in characterising the observable properties of voids and understanding their origin and dynamics \citep{hsw83,d+93,vdwvk93,sss94,colberg05,pbp06,vdwp11}. While a void can be defined in many ways \citep[see, e.g.,][and references therein]{colberg08}, the basic picture of a large, underdense, expanding region \citep{fg84,b85} has stood the test of time. Typical void sizes depend on the type of galaxy used to define them; e.g., in the main sample of the Sloan Digital Sky Survey, they can range from $\sim15\Mpc$ to $\sim30\Mpc$ \citep[e.g.,][]{Pan12,Sutter12}, while there are also examples of voids as large as $\sim100\Mpc$ \citep*[e.g.][]{GranettEtal2008}.

The presence of voids in galaxy surveys leads to many questions: whether galaxies that reside in void environments are special \citep{g+05,hrvb05}; whether large, deep voids are a challenge to structure formation in $\Lambda$CDM cosmologies \citep{b+92,hs10}, or whether they are a natural consequence of the well-understood dynamics of cold dark matter \citep{tc09}; whether voids can then be used as a cosmological tool to distinguish between models \citep{Ryden1995,pl07,lsd09,lw10,kvj09,baw10,dmnp11,LavauxWandelt2012,h+13,Melchior+13,Pisani13}; and whether their dynamics and statistics can be modelled analytically \citep{svdw04,fp06,daloisio2007}.

Analytical models for isolated voids have been well-studied in the literature for decades \citep{b85,b+92}. A major advance in their \emph{statistical} modelling was presented by \citet[][SvdW, in what follows]{svdw04}, who demonstrated that voids obey a hierarchy similar to that of halos. In particular, their analysis led to a prediction for the size distribution of voids based on the excursion set approach \citep{ps74,e83,bcek91}. Essentially, voids are modelled as regions that are initially underdense enough to reach shell-crossing by the present epoch. The SvdW analysis had three shortcomings, however; (a) it was based on an excursion set model using random walks in the smoothed density field with \emph{uncorrelated} rather than correlated steps, (b) it was entirely based upon the \emph{initial} or ``Lagrangian'' dark matter density, and (c), the intrinsic averaging of the excursion set walks, (on randomly selected position) was not taken into account. Recently, these shortcomings were overcome. In \citet*{pls12} the SvdW treatment was modified to account for both correlated steps in the random walks (which arise when using smoothing filters such as the real-space TopHat) as well as the fact that voids are identified in the evolved ``Eulerian'' field. In \cite{ARSC,AWWR}, it was shown that the consistency of the excursion set framework is preserved once the  barrier threshold is extended to stochastic modelling, and also shortcoming (a) was solved using an alternative path integral approach that we apply to voids in this work. 

Despite these improvements, excursion set void models cannot be directly compared with the distribution of observed galaxy voids. This is because these models are meant to describe voids in the dark matter, whereas the galaxies used to define voids observationally are biased tracers of dark matter. \citet{fp06} showed how galaxies can be included in the analysis by combining the SvdW excursion set calculation with the Halo Model \citep{ps00,Seljak2000}. As expected when using biased tracers, this increases the sizes of voids in a manner that is correlated with galaxy type (e.g., more luminous galaxies define larger voids on average). The size distributions of observed galaxy voids, e.g., those presented by \citet{Pan12} or \citet{Sutter12}, should therefore be compared with predictions such as those of \citet{fp06} and not with SvdW. 

Before doing this, however, it is important to ask whether the SvdW model \citep[or the improved version suggested by][]{pls12} gives a good description of voids identified in the dark matter density itself, which is possible in $N$-body simulations \citep{colberg05,jlh13}, and is one of the primary goals of this paper. \citet{jlh13} recently compared a modified version of the SvdW predictions (where the volume conservation is enforced) to the results of a void finder specifically built to identify spherical underdensities. Although it is plausible that this is the correct way of comparing the SvdW predictions with measurements, it ignores the highly aspherical, polyhedral shape that initial underdensities develop into as they form voids, and the unrealistic void volumes defined by the sharp-$k$ filter used in SvdW.  It is therefore interesting to ask whether the voids identified by popular algorithms (we use \zobov\ below) can be incorporated in an appropriate analytical framework that goes beyond the approximation of spherical evolution. Moreover, from a physical point of view, one also expects that voids tend to form near minima of the initial density field \citep[see][who demonstrated this in $N$-body simulations]{colberg05}, and it is then interesting to ask whether including a peaks constraint \citep{bbks86} in the excursion set calculation improves the comparison. 

The plan of the paper is as follows. In section~\ref{sec:analytical} we present analytical results for the void volume function based on path-integral calculations within the excursion set approach which we test against Monte Carlo simulations of random walks. We also present the results of including the peaks constraint in such a calculation, and extend both these results to the case of stochastic and scale-dependent void-formation thresholds. In section~\ref{sec:nbody} we turn to voids identified in $N$-body simulations. We describe the simulations and discuss the \zobov\ void finder. In particular, we explore the sensitivity of the latter to the choice of treatment of sub-structures within the identified voids, and compare our analytical results with the \zobov\ voids. We also study the effect of sampling the DM particles by considering voids identified with halos. The effect of the biasing is also tested by comparing this result with randomly selected DM particles.  In section~\ref{sdv} we check whether \zobov\ voids are consistent with the assumptions of the spherical evolution model by measuring the initial overdensity at an appropriately defined void center. In section~\ref{sec:su}, to test the sensitivity of our results to the particular void finder, we repeat some of our comparisons for voids found using a spherical-underdensity finder \citep[e.g.\ ][]{jlh13}. We conclude in section~\ref{sec:conclude} with a summary of our results and prospects for future work.

\section{The void hierarchy}
\label{sec:analytical}
If the statistics of voids carry cosmological information, then a successful theory should be able to predict void properties directly from the initial conditions once the cosmological background is known. The most naive idea is to link the site of a void to an underdense region in Lagrangian space. Assuming this initial depression evolves decoupled from the surrounding shear field, and is approximately spherical, then \cite{svdw04} have shown that the linear critical underdensity required to form a void at $z=0$ is approximately $\delta_v=-2.7$ in an Einstein-de Sitter universe. Unlike haloes, voids expand over time and repel matter. A spherical evolution model predicts that the Eulerian radius ($R_E$) of a void is $R_E\sim 1.7 R$, with $R$ its Lagrangian size. This deterministic mapping is more linear compared to the collapse of proto-halos (which contract by a factor $\sim 5.8$). The density within the void is $\Delta_v(z=0)\sim -0.8$. 

This rather simple analytical model is the building block which allows to pass from the statistical properties of voids in the matter density field to the statistical properties using biased tracers such as galaxies \citep{fp06}. The linear spherical threshold $\delta_v$ can be used to predict the site of void formation from the Lagrangian field. However, the dynamics of voids are subject to an additional, void-in-cloud effect (SvdW). This occurs in a region which is collapsing (or has collapsed) on a large scale $R_1$, but is underdense on a smaller scale $R_2$. In what follows, we make predictions for the void abundance using a realistic volume prediction within the standard excursion-set approach, and using a modified peak-excursion set approach. For both cases we also extend the spherical threshold to more general class of barriers defined by a Gaussian and a Log-normal distribution.

\subsection{Excursion set approach}
The standard excursion-set theory \citep{bcek91} is a useful framework to compute the abundance of dark matter halos, and can also be applied to voids. The key assumption is to equate the volume fraction in voids of radius $R$ to an appropriate first-crossing distribution:
\begin{equation}
V\frac{\der n}{\der\ln R}=f(\sigma) \left|\frac{\der\ln\sigma}{\der\ln R}\right|
\label{volfunc}
\end{equation}
where $f(\sigma)\equiv 2\sigma^2 \mathcal{F}(\sigma)$ is the so-called multiplicity function and $\mathcal{F}\equiv dF/dS$ is the derivative of the volume fraction arising from the first-crossing problem. 
% f(sig) = 2sF(s) = usually denoted nu f(nu)
Let us denote the probability density that an overdensity smoothed on a scale $R(S)$ is below a critical threshold $B$ by $\Pi(\delta,S(R))$. We have 
\begin{equation} 
F(S(R))=-\int_{-\infty}^{B} \Pi(\delta,S(R)) \; d\delta+C,
\end{equation}
where $C$ is a constant independent of the scale $R$, and $S$ is the variance of the associated field:
\begin{equation}
S\equiv\langle\delta^2(R)\rangle\equiv\sigma^2=\frac{1}{2\pi^2}\int dk\,k^2P(k)\tilde{W}^2(k,R)
\label{variance}
\end{equation}
Once the filter $W(k,R)$ is specified, there is a one-to-one mapping between the scale $R$ and the variance $S$.

This formalism can be extended in the void hierarchy where voids are characterized by their volume rather than the mass they encapsulate.

Another common quality of the excursion-set theory is that the so-called cloud-in-cloud issue is solved: a collapsing structure cannot be embedded in a larger one which would lead to miscounting the number of halos. This issue appears if the smoothed over-density crosses the threshold at multiple smoothing scales and can be treated by adding an absorbing boundary condition: $\Pi(\delta=B,S)=0$ such that the largest scale defines the mass $M(R)$. In the case of voids, the void-in-cloud process, describing collapsing voids, is important to take into account, as described in SvdW for a sharp-$k$ filter (SK). Therefore we must distinguish between the barrier associated with halos (denoted $B_h$) and the one associated with voids (denoted $B_v$). Thus all the game is to compute the $\Pi(\delta,S)$ under the condition that $\Pi(\delta,S=0)=\delta_D(\delta)$, $\Pi(\delta=B_h,S)=\Pi(\delta=B_v)=0$ and compute the first-crossing $F(S|\delta(S')<B_h(S'))$ with $S'<S$. For a sharp-$k$ filter and a constant barrier (e.g., $B_v=-2.7$, $B_h=1.686$) the solution of this system is given in SvdW. The extension to a linear moving barrier of the same slope ($B_v=\delta_v-\beta S$, $B_h=\delta_c-\beta S$) can be found in appendix C of SvdW, while the extension to positive slope has been worked out in Appendix A of \cite{fp06}. Note that for the halo barrier, ellipsoidal collapse predicts a positive slope. However, we will see in $\S$\ref{sdv} that for the void threshold, it seems that a negative slope is in better agreement with the Lagrangian barrier. However, before jumping to the barrier criteria we should emphasize that all those predictions hold for a particular type of filter, a TopHat in Fourier space (sharp-$k$, SK). The volume encapsulated by such a filter is given by\footnote{see also discussion after Eq.\ (36) in \cite{MR1}.}:
\begin{equation}
V_{SK}(R)=\int d^3R\; W_{SK}(R)=6\pi^2 R^3 -12\pi R^3 \int_{0}^{\infty} \cos x dx.
\end{equation}
One could argue that the divergent integral part can be set to zero \citep[see, e.g.,][]{LaceyCole}. However, it is more difficult to picture the shape associated with such a void and in addition, this volume is never used in observations or in $N$-body simulations to define structures. Therefore, if we assume a spherical-shell evolution of the void, for consistency, the appropriate filter should be a TopHat in real space (SX filter), which defines a spherical volume. However, in this case there is no exact analytical solution to the first crossing. One could run Monte Carlo walks \citep{bcek91} and solve the exact associated first-crossing, which would be straightforward \citep{pls12}. Nevertheless, a path-integral approach \citep{MR1,CA1} can be used to compute analytically the correction induced by such a filter (see \cite{ms12} for alternative methods). The SX filter introduces small corrections to the SK case, which can be computed perturbatively and applied to halo formation in the excursion set framework. The amplitude of this correction is weakly dependent on the smoothing scale, and is set by the linear matter power spectrum. This method has been shown to be very accurate and to converge well: the exact Monte Carlo solution matches the analytical approximation with high accuracy \citep{CA2,ARSC}. In what follows, we investigate the pertinence of the void-in-cloud effect for realistic halo thresholds in the context of the excursion set theory, and show that a one-barrier threshold is a very good approximation to the exact Monte Carlo solution, providing a simple analytical formula for the SK filter. Finally we extend this prediction to the SX filter. 
Our results are consistent with previous work by SvdW for sharp-$k$ filtering and \citet{pls12} for (SX) filtering. See also \cite{ZhangHui2006,LamSheth2009} for a complementary approach.

\medskip

\cite{ARSC} found that within the excursion set framework, any consistent barrier should have an intrinsic scatter due to the randomness of the position that the excursion-set theory assumes in order to compute the fraction of collapsed regions. Note that deviations from spherical collapse also contribute to this scatter\footnote{The reconstruction of the barrier for the center of mass (ie: on the peak of the proto-halo) also shows a scatter \citep{ARSC,Roberton,AWWR}} . Over the range they investigate, they found that a Gaussian barrier with a mean value of $\langle B \rangle(S)=\delta_c+\beta S$ and r.m.s. $\sqrt{D_B S}$ is consistent with the initial Lagrangian critical overdensity leading to halo formation, and predicts a mass function which is in very good agreement with $N$-body simulations \citep[e.g.,][]{AC1,CA2}. Therefore, in order to test the void-in-cloud effect on the abundance of void, we assume a realistic barrier for halo and void formation (a diffusive drifting barrier). Note that in this model the random walk performed by the barrier is not correlated with the one performed in $\delta$. See \cite{ARSC} for a discussion on this assumption.

Following \cite{bcek91}, we perform Monte-Carlo random walks to solve the first-crossing associated with a generic filter and barrier, and we implement the condition that walks which cross the void barrier on a scale $S_1$ never cross the halo barrier on a smaller scale $S_2<S_1$. For the halo barrier, we take $\beta=0.1,D_B=0.4$. Similarly, for the void barrier we consider a barrier with a Gaussian distribution characterised by a mean $\langle B_v \rangle=\delta_v-\beta S$ and r.m.s. $\sqrt{D_B S}$, with the same choices $\beta=0.1, D_B=0.4$. Note that this choice of parameters is rather arbitrary, because there is no theoretical prediction for them. We flip the sign of the slope for reasons that will become clear in $\S$\ref{sdv}. 

\begin{figure*}
\begin{center}
\includegraphics[scale=0.42]{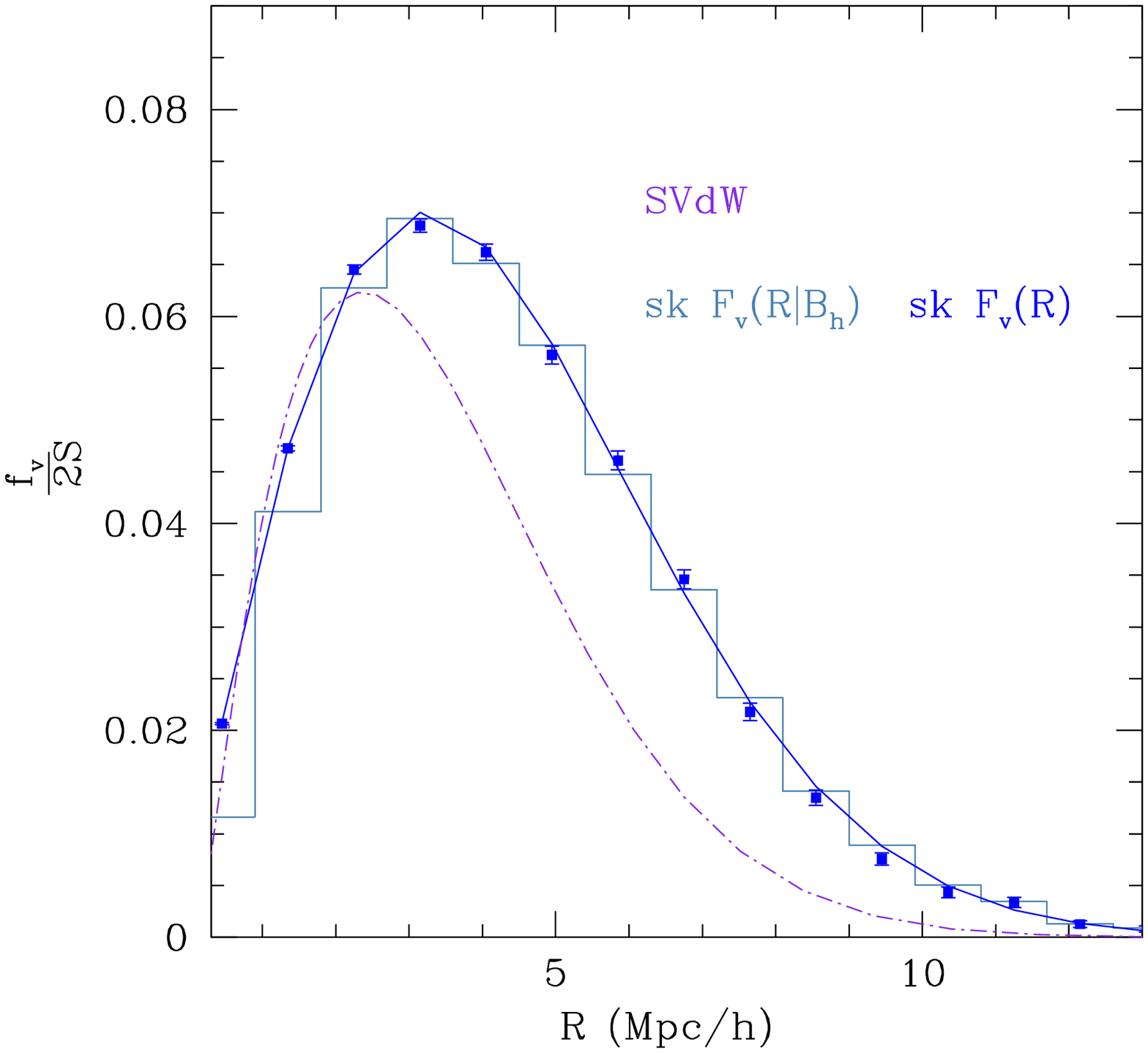}
\includegraphics[scale=0.42]{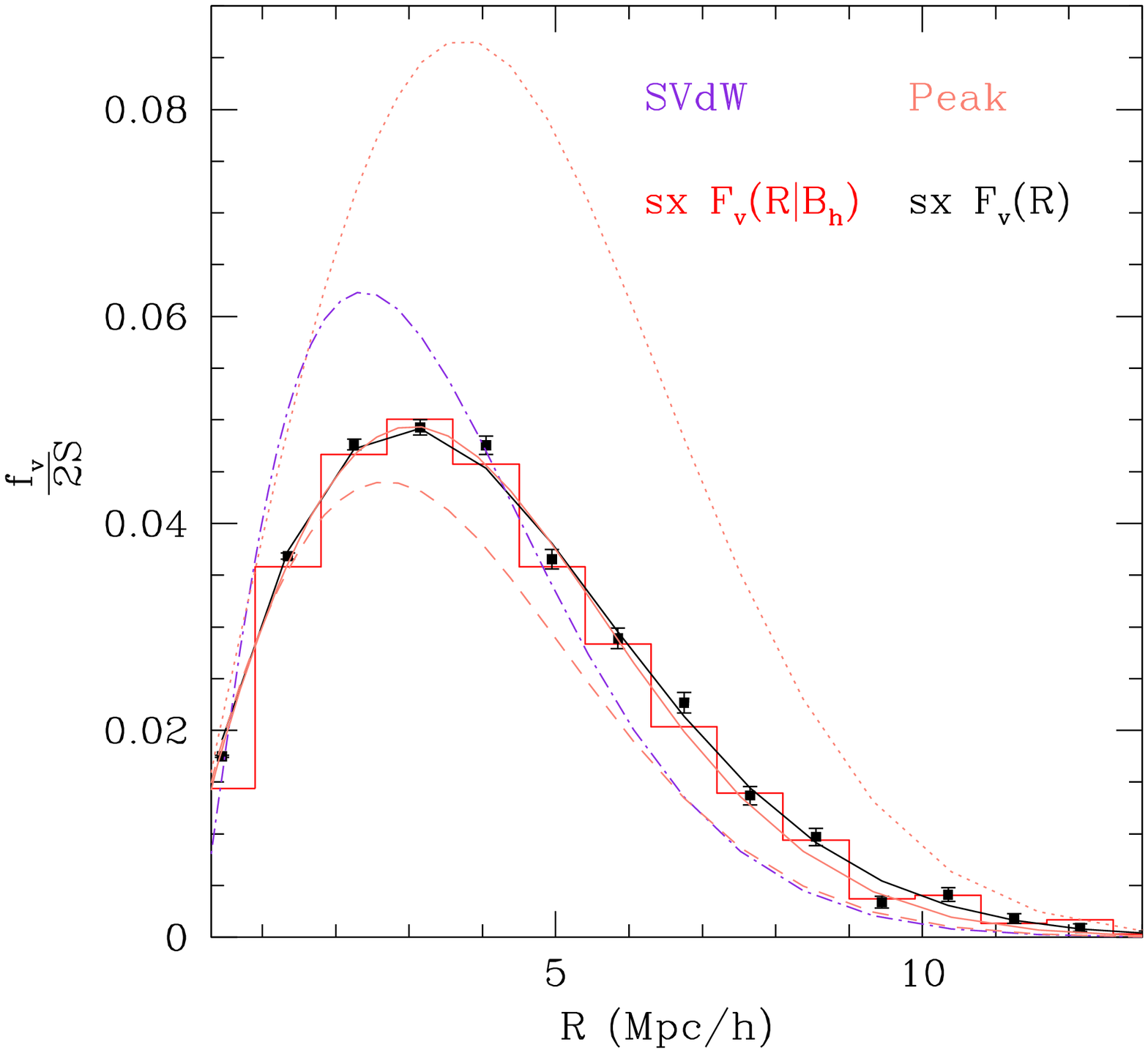}
\caption{Void multiplicity function as a function of void radius for Markovian random walks (SK) (left panel) and correlated walks (SX) (right panel). The histogram shows the exact solution including the treatment of the void-in-cloud effect while the dots neglect this issue. Black and blue solid lines are the theoretical predictions of Eq.\ (\ref{fsigma0},\ref{ftot}), orange lines correspond to peak prediction (see text) and the purple dot-dashed line shows the SvdW reference (SK). }\label{fig1}
\end{center}
\end{figure*}

The Monte Carlo results for the SK filter are shown in Fig.\ \ref{fig1}. The light-blue histogram shows the Monte-Carlo result associated with the two-barrier condition, while the blue dots neglect the void-in-cloud effect. As we can see, the void-in-cloud effect operates at low radius. This effect also depends on the halo threshold as it was discussed in \cite{svdw04}. For a drifting diffusive barrier and Markovian walks (SK filter), the void-in-cloud effect appears at $R<3\ {\rm Mpc}/h$. The blue solid line is the analytical prediction of the excursion-set theory for a diffusive drifting void threshold with $\langle B_v(S)\rangle=\delta_v-\beta S$ and $\langle B_v(S_1)B_v(S_2)\rangle=D_B\rm{min}(S_1,S_2)$. This solution neglects the void-in-cloud effect and is exact for the sharp-$k$ filter.  It was computed as in \cite{CA1,CA2}, leading to a Markovian multiplicity function (SK) for voids:
\begin{equation}
 f_0(\sigma)=\frac{\delta_v}{\sigma}\sqrt{\frac{2a}{\pi}}\,e^{-\frac{a}{2\sigma^2}(\delta_v-\beta\sigma^2)^2},\label{fsigma0}
\end{equation}
with $a=1/(1+D_B)$. 

The original prediction of \cite{svdw04} (SvdW) is shown in purple. Note also that Eq.(\ref{fsigma0}) reproduces with high accuracy SvdW by simply setting $\beta=0,D_B=0$ ($R>2\ {\rm Mpc}/h$). However, in order to have a coherent volume definition, we should consider walks smoothed with a (SX) filter when computing the multiplicity function. In this case, we use the same path integral technique as \cite{MR1,MR2,CA2,AC2}.  Taking into account void-in-void and neglecting void-in-cloud effects, we find that the non-Markovian corrections for a diffusive drifting barrier are:
\begin{align}
f_{1,\beta=0}^{m-m}(\sigma)&=\tilde{\kappa}\dfrac{\delta_v}{\sigma}\sqrt{\frac{2a}{\pi}}\left[e^{-\frac{a \delta_v^2}{2\sigma^2}}-\frac{1}{2} \Gamma\left(0,\frac{a\delta_v^2}{2\sigma^2}\right)\right]\,,\\
f_{1,\beta^{(1)}}^{m-m}(\sigma)&=a\,\delta_v\,\beta\left[\tilde{\kappa}\,\text{Erfc}\left( \delta_v\sqrt{\frac{a}{2\sigma^2}}\right)+ f_{1,\beta=0}^{m-m}(\sigma)\right]\,,
\label{beta2}\\
f_{1,\beta^{(2)}}^{m-m}(\sigma)&=-a\,\beta\left[\frac{\beta}{2} \sigma^2 f_{1,\beta=0}^{m-m}(\sigma)-\delta_v \,f_{1,\beta^{(1)}}^{m-m}(\sigma)\right],
\end{align}
where $\tilde{\kappa}=a\,\kappa$, and $\kappa$ is set by the linear matter power spectrum. For a vanilla $\Lambda$CDM universe, $\kappa \sim 0.465$, giving the following void total mutliplicity function for a sharp-$x$ filter: 
\begin{equation}
f_v(\sigma)=f_0(\sigma)+f_{1,\beta=0}^{m-m}(\sigma)+f_{1,\beta^{(1)}}^{m-m}(\sigma)+f_{1,\beta^{(2)}}^{m-m}(\sigma)\label{ftot}
\end{equation}
To test this prediction, we show also in Fig.\ \ref{fig1} the exact Monte-Carlo solution associated with the same barrier as before, including the void-in-cloud effect (red histogram), neglecting the void-in-cloud effect (black dotted), and the theoretical prediction of Eq.\ \ref{ftot} (black solid line). As we can see, the agreement with the exact solution is quite accurate over the all range in radius. Note also that the correlations between steps induced by the SX filter decreases the number of voids by a non-negligible factor; thus, the effect is important and should be properly implemented in any void-abundance prediction. Furthermore, the difference between the two-barriers model and the solution which neglects the void-in-cloud process is less important than for the sharp-$k$ case, a point first made by \cite{pls12}. Similarly  to the halos cloud-in-cloud process, the physical reason why the void-in-void or void-in-cloud effects influence only small scales is that large-scale voids are most likely to be at the top of the hierarchy, not embedded in even larger voids or haloes.  For SX filters, this Monte Carlo shows that the probability that an initial underdense patch of matter with Lagrangian radius $>1\ {\rm Mpc}/h$ is embedded in an overdense larger region is negligible. These results are also in agreement with \cite{jlh13}. Note that the small influence of the void-in-cloud effect could also be due to the drifting terms which effectively increase the separation between the halo and void barriers. Overall those Monte Carlo tests show that Eq.\ (\ref{ftot}) is a good prediction for the void abundance as long as the excursion set assumptions (e.g.\ averaging the smoothed field over random positions) can be applied to the description of void statistics and the void-in-cloud process is negligible. We describe an alternative peaks approach in the next section.

\subsection{Peaks approach}
In addition to their $2$-barrier sharp-$k$ random-walk model, SvdW also discussed alternative models based on counting density minima in the initial conditions. The model that they called ``adaptive troughs'', which was based on previous work by \citet{aj90} for the halo mass function, is especially interesting for us, because recent work on the nature of random walks with correlated steps sheds new light on its interpretation.

The adaptive-troughs model states that the void multiplicity function can be written by using the \citet[][BBKS in what follows]{bbks86} result for counting density peaks/troughs and including the effect of a variable smoothing filter:
\be
f(\sig) = \frac{{\rm e}^{-\del_v^2/2\sig^2}}{\sqrt{2\pi}\gam}
                          \frac{V}{V_\ast}\, \int_{0}^\infty\der x\, xF(x)\,p_{\rm G}(x-\gam|\del_v|/\sig;1-\gam^2)\,.
\label{vfv-troughs}
\ee
Here $V=4\pi R^3/3$ is the Lagrangian volume of the void, $p_{\rm G}(y-\mu;\Sigma^2)$ is a Gaussian in the variable $y$ with mean $\mu$ and variance $\Sigma^2$, and \gam\ and $V_\ast$ are ratios of spectral integrals that appear when counting density peaks/troughs,
\be
\gam\equiv \sig_1^2/(\sig_0\sig_2)\quad;\quad V_\ast \equiv (6\pi)^{3/2}\sig_1^3/\sig_2^3\,,
\label{gam-Vst}
\ee
where
\be
 \sig_j^2 = \int \frac{\der^3k}{(2\pi)^3}\,P(k)\,k^{2j}\,{\rm e}^{-k^2R_{\rm G}^2}.
\label{sigma-j}
\ee
The Gaussian smoothing scale $R_{\rm G}$ depends approximately linearly on the Lagrangian radius $R$ and is discussed below.

The integral in \eqn{vfv-troughs} is over the peak curvature $x=-\nabla^2\del/\sig_2$, and involves the weighting function $F(x)$ given by
\begin{align}
F(x)&=\frac12\left(x^3-3x\right)\left\{\erf{x\sqrt{\frac52}}+\erf{x\sqrt{\frac58}}
  \right\} \notag\\
&\ph{x^3-3x}
+ \sqrt{\frac2{5\pi}}\bigg[\left(\frac{31x^2}{4}+\frac85\right){\rm
    e}^{-5x^2/8} \notag\\
&\ph{\sqrt{x^3-3x+\frac2{5\pi}}[]}
+ \left(\frac{x^2}{2}-\frac85\right){\rm
    e}^{-5x^2/2}\bigg]\,,
\label{eqn-bbks-Fx}
\end{align}
which is the result of integrating over peak shapes (equations~A14--A19 in BBKS). While there is no closed form expression for the multiplicity \eqref{vfv-troughs}, the integral involved is straightforward to compute numerically, and we also note that BBKS provide a very accurate analytical approximation in their equations~(4.4, 4.5, 6.13, 6.14).

Recently, \citet{ps12} pointed out, based on results obtained by \citet{ms12}, that the multiplicity in \eqn{vfv-troughs} is an excellent approximation to the first-crossing distribution of the constant barrier $B=\del_v$ by \emph{peak-centered} random walks with correlated steps. Moreover, as argued by \citet{pls12}, accounting for the complications introduced by the fact that voids are identified in Eulerian rather than Lagrangian space does not lead to significant effects when dealing with walks that have correlated steps. In particular, \citet{pls12} showed (see their Figure 3) that the appropriate first-crossing distribution for Eulerian voids (under the assumption of spherical evolution) is indistinguishable from that of a single constant barrier of height $\del_v$ for all but the smallest voids. In other words, taken together, the results of \citet{pls12} and \citet{ps12} suggest that \eqn{vfv-troughs} should be a good model of void abundance, if one expects voids to have formed near initial density minima.

There is a technical issue related to the choice of Gaussian filtering with scale $R_{\rm G}$ in defining the spectral integrals in \eqn{sigma-j}. Ideally one would use TopHat (SX) filtering to define these integrals. However, in this case the identification of peaks for the CDM power spectrum becomes ill-defined since, e.g., $\sig_2$ is no longer well-defined. Gaussian filtering avoids this problem, and all results in BBKS assume this. In order to make the calculation consistent with the standard assumption of defining \del\ using TopHat filtering, \citet*{psd13} proposed the following: to identify peaks/troughs, one can use spatial derivatives of the Gaussian-filtered density contrast $\del_{\rm G}(R_{\rm G})$ so that $\sig_1^2=\avg{(\nabla\del_{\rm G})^2}$ and $\sig_2^2=\avg{(-\nabla^2\del_{\rm G})^2}$ are well-defined. The \emph{heights} of these density extrema, on the other hand, can be defined using the TopHat-filtered $\del_{\rm TH}(R)$. The connection between the two smoothing scales $R_{\rm G}$ and $R$ follows by demanding $\avg{\del_{\rm G}|\del_{\rm TH}} = \del_{\rm TH}$. Since $\del_{\rm G}$ and $\del_{\rm TH}$ are both Gaussian distributed, this amounts to requiring $\avg{\del_{\rm G}\del_{\rm TH}}=\avg{\del_{\rm TH}^2}=\sig^2$. This can be solved numerically and, in practice, gives $R_{\rm G}\approx0.46R$ with a slow variation. To be fully consistent, one must also redefine \gam\ as
\begin{align}
\gam\to\gam_{\rm m} &= \sig_{\rm 1m}^2/(\sig\sig_2) \notag\\
&= \frac1{\sig\sig_2}\int\frac{\der^3k}{(2\pi)^3}\,P(k)\,k^{2}\,{\rm e}^{-k^2R_{\rm G}^2/2}\tilde W(kR)
\label{gam-mix}
\end{align}
with $\tilde W(kR)$ the Fourier transform of the TopHat filter, and where \sig\ is defined in \eqn{variance}. (In practice, the TopHat-filtered $\sig^2$ and Gaussian-filtered $\sig_0^2$ differ by at most $\sim5\%$ or so.)

These results can also be extended to the case when the barrier relevant for void formation is stochastic and/or scale-dependent. Following \citet{psd13}, for a barrier of the form
\be
B_v = \del_v - \beta_{\rm pk}\sqrt{S}
\label{esp-stochbar}
\ee
with the slope $\beta_{\rm pk}$ a stochastic quantity with distribution $p(\beta_{\rm pk})$ in general, the void multiplicity becomes
\begin{align}
f(\sig) &= \int\der\beta_{\rm pk}\, p(\beta_{\rm pk})\, \frac{{\rm e}^{-(\del_v-\beta_{\rm pk}\sig)^2/2\sig^2}}{\sqrt{2\pi}\gam_{\rm m}}
\frac{V}{V_\ast}\, \notag\\
&\ph{V}\times
\int_{\beta_{\rm pk}\gam_{\rm m}}^\infty\der x\, (x-\beta_{\rm pk}\gam_{\rm m})F(x)\,\notag\\
&\ph{V/V_\ast}\times
p_{\rm G}(x-\beta_{\rm pk}\gam_{\rm m}-\gam_{\rm m}|\del_v|/\sig;1-\gam_{\rm m}^2)\,.
\label{vfv-ESPstoch}
\end{align}
The resulting multiplicity is shown in Figure~\ref{fig1} for different parameters. The dotted orange curve shows the prediction from \eqn{vfv-troughs} for the constant barrier $B_v=\del_v=-2.7$. The dashed orange curve shows the effect of introducing a negative drift with constant slope $\beta_{\rm pk}=0.5$ (formally, equation~\ref{vfv-ESPstoch} with $p(\beta_{\rm pk})=\dir(\beta_{\rm pk}-0.5)$), while the orange solid curve shows \eqn{vfv-ESPstoch} setting $p(\beta_{\rm pk})$ to be lognormal with mean $0.5$ and variance $0.25$. We return to those choices of parameters later.

First of all observe that the standard peak-based ``adaptive troughs'' prediction (orange dotted curve) leads to a higher amplitude for the abundance of voids compared to the standard excursion set approach with a diffusive drifting barrier (see also SvdW). 
In addition, introducing scatter in the barrier height in the peaks prediction also tends to increase the number of voids, but it is a subdominant effect compared to the negative drift which decreases the amplitude. The solid orange line implements both scatter and drift while the orange dashed line neglects the scatter.
 Finally we note that interestingly, both predictions for stochastic barriers with a negative drift are close to each other. 
For comparison, we also show the original SvdW prediction as the dot-dashed curve in which the relation between the density variance $S$ and Lagrangian radius $R$ was computed using the SX filter. While this is the usual manner in which the SvdW result is used, we emphasize that doing so is technically inconsistent, since the derivation in SvdW assumed SK filtering. Note also that the effect of the SX filter on the void abundance is to decrease the total number of voids.\\

Before moving to the next section, we should mention that the cumulative void volume fraction $\mathcal{F}(R>R_{\rm min})$ can be computed from Eqs.\ (\ref{vfv-ESPstoch}) and (\ref{ftot}) only in the regime where the void-in-cloud process is negligible (i.e.: $R_{\rm min} >1 {\rm Mpc}/h$).

\section{Comparison with N-body simulations}
\label{sec:nbody}
In order to test our theoretical predictions, we measure void abundances from $N$-body simulations using the \zobov\ void finder \citep{Neyrinck2008}, described in the next subsection.  Firstly we consider the dark-matter field of the DEUS $N$-body simulations\footnote{www.deus-consortium.org}, described in \cite{Alimi2010,Courtin2011,yann}, without any particle subsampling. We use two box sizes, of length $162$ and $648\ {\rm Mpc}/h$, both with $1024^3$ particles, realized using the RAMSES code \citep{Ramses} for a $\Lambda$CDM model calibrated to WMAP 5-year parameters $(\Omega_{\rm m},\sigma_8,n_s,h,\Omega_{\rm b})=(0.26,0.79,0.96,0.7,0.0456)$. Secondly, we study the effect of subsampling the particles and using biased tracers to identify voids. For this purpose we use randomly selected DM particles equal to the total number of halos in the simulations. To study the effect of the bias, we compare the resulting void function to the one obtained using the halo catalogues as tracers. In both boxes, halos are identified using the friends-of-friends algorithm with linking length b=0.2.

\subsection{ZOBOV}
\label{sec:zobov}
The \zobov\ (ZOnes Bordering On Voidness) void finder \citep[][N08]{NeyrinckEtal2005,Neyrinck2008} is designed to be parameter-free. \zobov\ uses the adaptive, parameter-free Voronoi tessellation to estimate the density \citep[e.g.][]{SchaapVandeWeygaert2000} at every particle. A void is grown around each local-density-minimum particle using a watershed transform \citep[e.g.][]{PlatenEtal2007}: in an analogy to rain falling and flowing across a terrain, a particle $p$ gets associated with a density minimum $p_{\rm min}$ if a particle-to-particle path on the tessellation down the steepest density gradients from $p$ ends at  $p_{\rm min}$.  

\zobov\ forms a parameter-free partition of all particles into so-called `zones,' each of which is a watershed-region flowing down into a single density minimum.  It then returns a void catalogue, consisting of voids (sets of zones joined together) and subvoids. But we introduce two parameters to prune the raw void catalogue to something physically corresponding to our theoretical model, in which there are no subvoids; i.e.\ `voids in voids' are not double-counted.

To obtain a disjoint set of voids with boundaries that are likely not spurious, we apply the `specifying a significance level' strategy described in N08 to the raw void catalog. In this strategy, a boundary between two adjacent zones is declared to be real if the `density-contrast ratio,' i.e.\ the ratio between the ridge density (the lowest-density along the ridge separating the voids) and the density minima, exceeds a threshold corresponding to a two-sigma (95\%) probability that a void did not arise from Poisson noise (N08).

 We also introduce a threshold to the minimum density in each void, to eliminate local density minima in high-density regions (which will occur by chance in a sufficiently well-sampled high-density structure). 
As shown by N08, removing voids with density contrast under the two-sigma threshold will typically also remove these high-density voids, but to be sure about this, we apply a threshold at the minimum density found in the void, called the `core density' (void with minimum density).

In theory, a spherical void has $\Delta_v=-0.8$. Using \zobov\, a void is composed of several zones which have different densities. The total mean density of all zones is what we expect to correspond best to the critical $\Delta_v=-0.8$. However, several zones can have much higher density than the core zone.  Thus, we use values of the core threshold  $\le -0.8$.

%We also do not allow subvoids to belong to deeper voids if the lowest density on the ridge between the zones exceeds a parameter, $\Delta_{\rm ridge}$, that for simplicity we set to be the same as $\deltavmin$, discussed below.

Unfortunately it is difficult to know {\it a priori} what value to use for this threshold. It can be calibrated by measuring abundances of density contrast ratios in Poisson point samples, but this describes its statistical, not physical, significance. If this density-contrast ratio is used to judge voidness, the void catalogue best corresponding to a physical set of voids would likely differ based on the mass resolution, or sampling level. In what follows, we use both $\Delta_{v}^{\rm{min}}=-0.9$ and $\Delta_{v}^{\rm{min}}=-0.8$ which is one common choice used in the literature (e.g. \citet{LavauxWandelt2012}, \citet{Pisani13}, \citet{Chanetal2014}) .

One might wonder why we do not cut directly on the (volume-weighted) average density within the void. This average density is easily computed from quantities in the void catalogue, but in fact it can be quite noisy. This is because a watershed transform does not give boundaries that necessarily correspond to density contours; all that is required for a particle to belong to a void is that the particle is up a steepest density gradient from a density minimum, so in fact haloes might be included at the edge of a void.  Still, a void's average density from \zobov\ gives additional information about it, and we will use it below.

We use a non-periodic box cut from a larger (periodic) volume, which approximates the situation one might consider observationally. To deal with boundaries, we follow the typical approach done with the \zobov\ algorithm \citep{GranettEtal2008,Sutter12}: we surround the box with a dense set of border particles. In the void-finding step, we exclude any particle that has a border particle detected as a neighbor. Unlike \citep{Sutter12}, however, we do not remove voids from the catalogue that could be rotated to intersect the boundary.

\subsection{The abundance of voids in the DM density field and for biased tracers}

The number of voids identified with \zobov\ is sensitive to the density of the tracers inside the box. Naturally, if the number density of particles is high, then smaller voids will be detected. \zobov\ is designed to have low sensitivity to the sampling level for large voids, but the boundaries of voids change slightly when adding or subtracting particles randomly, so large voids are sometimes not exactly preserved. Also, increasing the mass resolution in a CDM simulation adds small-scale power, so decreasing mass resolution is not necessarily the same as randomly removing particles.  
As a result of these issues, the void abundance function can change slightly when particles are subsampled. In \citep{Pisani13,Chanetal2014}, the authors subsampled particles of the DM field to match the density of the SDSS survey. In this work, we present two extreme cases: we consider the full DM density field and a sample of it for which the DM density equals the density of halos in the simulation.

 \begin{figure}
\begin{center}
\includegraphics[scale=0.45]{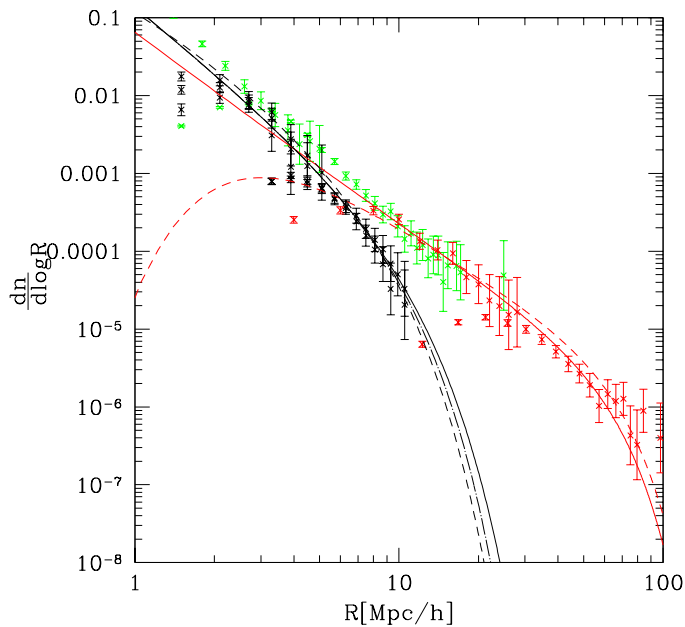}
\caption{The void size distribution at $z=0$ measured in simulations with different density tracers and  choices of post-processing. The black and green crosses are from the unsampled dark matter density fields and follow respectively case 1 and case 2 for the choice of post-processing (see text). The red crosses are from the halos and follow case 2 for the choice of post-processing. Overplotted are the theoretical predictions of SvdW (dashed curves), DDB (solid curves) and peaks approach (dot-dashed curve). }\label{fig2}
\end{center}
\end{figure}

For the full DM density field, we analyzed three sub-cubes of size $(40.5\ {\rm Mpc}/h)^3$ from the $(162\ {\rm Mpc}/h)^3$ box and two sub-cubes of size $(162\ {\rm Mpc}/h)^3$ from the $(648\ {\rm Mpc}/h)^3$ box. The total number of dark matter particles within the $40.5\Mpc$ and the $162\Mpc$  sub-boxes is $N \sim 1.6\times 10^7$, which is about the limit the qhull algorithm (external module of \zobov) can treat. \\

%This is the largest plausible choice, since again the spherical-expansion expectation has $\Delta=-0.8$ for the mean density in the void, which could substantially exceed $-0.8$. This analysis does not comport any additional post-processing. As a result, this void catalogue contains subvoids and fake-voids (e.g. with arbitrary density contrast).  \\

To avoid counting the same structures twice, we remove subvoids by declaring boundaries to be real at a density contrast corresponding to a two-sigma cut in a Poisson realization. This post-processing also removes 
spurious voids or sub-voids arising from Poisson fluctuations in high-sampling limit. We used a two sigma cut as a compromise between having an accurate sample of disjoint voids (without subvoids) and a large number of voids. If we reduce the cut to one-sigma, this would declare shallower boundaries separating subvoids within voids to be real, cutting up larger voids and shifting the distribution to small radius. Likewise, increasing the threshold removes walls between voids, which would increase the number of larger voids. We also try setting \zobov's core-density threshold equal to $-0.9$. We define this choice of post-processing as case 1 (disregarding void boundaries under a 2-sigma significance cut, and using a $-0.9$ minimum-density threshold). The corresponding results for void abundances are shown in Fig.\ \ref{fig2} in black crosses.
First three sub-cubes probe voids with radii $R\sim 1.5-5 \rm Mpc/h$ while the other two sub-cubes probe voids with $R\sim 4.5-10 \rm Mpc/h$.

 Our second choice of post-processing is to put a minimum-density threshold of $-0.8$ without a void boundary cut. We define this fiducial model as case 2. The result for case 2 on the unsampled dark matter density field is shown by the green crosses in Fig.\ref{fig2} for two sub-cubes of size $(40.5\ {\rm Mpc}/h)^{3}$ and $(162\ {\rm Mpc}/h)^{3}$ . In this case, the smaller sub-cube probes voids with radii $R\sim 1.5-5\rm Mpc/h$, while the larger sub-cube probes voids with $R\sim 4-18 \rm Mpc/h$.

Note that in all cases, we removed from the voids list (output files of \zobov), those which have a large {VoidDensContrast} ($>10$). We further put a cut in the void function when the Poisson noise is higher than 50$\%$, for clarity of the Figure. For both case 1 and case 2 we find a convergence of the void abundance within the different sub-boxes with different mass resolutions. Sensitivity to the choice of post-processing is significant for high density tracers. This is because the voids from a high density sample will contain a lot more subvoids and spurious voids compared to voids identified in low-density sample.

Unsurprisingly, using a two-sigma cut of post-processing (case 1) reduces significantly the abundance of voids. The choice of core-density threshold, $-0.9$, is rather arbitrary and is a compromise between matching the theory lines and choosing a threshold closer to what we can expect from a void with under-density $\Delta=-0.8$ (a cut in the minimum density of a void should be lower than $-0.8$ to achieve a mean density in the void of $-0.8$).

To map the Lagrangian theory of Eq.\ref{volfunc} to effective void radius ($R=(3 V/4\pi)^{1/3}$, where $V$ is the void volume reported by \zobov), we use the spherical model and set $1.7 R_{Lag}=R$. 
One could adopt a different approach and use a different mapping which might arise from aspherical voids consistent with the Lagrangian underdensity (linear void threshold). We do not investigate this issue here. In the case of halos, this issue is not very relevant. In fact, spherical overdensity halo-finders are based on the non-linear spherical collapse model, although the linearly extrapolated spherical-collapse threshold does not work in detail for predicting the halo mass function. Therefore we adopt the same pragmatic approach for voids. 
The black dashed curve in Fig.\ref{fig2} shows the SvdW prediction with the usual $\delta_v=-2.7$ and $\delta_c=1.68$. The black solid curve use Eq.\ (\ref{ftot}), while the black dot-dashed curve use Eq.\ (\ref{vfv-ESPstoch}), with the same parameters as in Fig.\ \ref{fig1}. 

All the black theory curves match the data reasonably well. The parameters for the barrier we adopt were, strictly speaking, motivated by parameters for the halo, not void, mass function (\cite{ARSC,psd13}). We do not expect the same criteria to hold for both, although this is a good starting point. The agreement is a nice result, since Eqs.\ (\ref{ftot},\ref{vfv-ESPstoch}) correspond to the prediction of the SX filter, which assumes a spherical volume to map the variance to the radius of the voids; this is one of the main differences compared to the SvdW prediction. Note that, in contrast to the voids identified by \citet{jlh13}, the SvdW curve is quite close to the volume function of our \zobov\ voids, with no factor $\sim5$ offset. In fact, re-defining voids as proposed by \cite{jlh13}, might be more consistent with the spherical shell approximation. In order to defined voids as `spherical underdensities.' \citet{jlh13} use density minima found by \zobov\ to get initial void and subvoid centers, but then they define the boundaries quite differently: for each void and subvoid, they start from a large radius about the center, and decrease the radius until the underdensity reaches $\Delta_v = -0.8$.  While this procedure intuitively corresponds well to a spherical-shell approximation, the voids can be quite aspherical, so this procedure is not guaranteed to give the optimal correspondence with theory. In full nonlinearity, voids run into each other and their boundaries typically depart substantially from a spherical shape; these arbitrary shapes will be picked up by \zobov.  Note also that \citet{jlh13} compare the `spherical-underdensity' voids with the excursion-set theory associated with the SK filter. 

\medskip

Finally we investigate the effect of sampling the tracers to identified voids. In fact, for very low-density tracers, we expect void catalogues to contain large voids, while for very dense tracers (as for the full DM density field at the resolution here), voids are rather small. Therefore we probe large voids by running \zobov\ on the halo positions of our two simulations with a core-density threshold $-0.8$ and no sigma cut (case 2).  The halo densities in our two boxes are $\sim 0.073 h^{3}/Mpc^{3}$ and $\sim 0.001 h^3/Mpc^{3}$ for respectively the $162,648 Mpc/h$ box sizes. The result is shown in Fig.\ref{fig2} by the red crosses.  The $162 Mpc/h$ simulation probes void between $\sim 10-20 \rm Mpc/h$ while the $648 \rm Mpc/h$ simulation probes voids between $\sim 20-80 \rm Mpc/h$. Again, we observe a good overlap of the void abundance between the different simulations.

In this low density sample, our choice of post-processing would not change significantly the abundance of voids. In \citep{Chanetal2014} the authors observed that if the density of the tracers is not high enough, smaller voids tend to artificially merge to form larger voids. To avoid this effect, a statistical threshold like the one we chose for the unsampled dark matter particle is not enough. Nevertheless, as we can see in Fig.\ref{fig2}, the abundance of these voids can be fitted by adjusting the value of $\delta_c$ which enters into the theoretical prediction. The red dashed curve shows the theory of SvdW using $\delta_v=-0.6$.  The red solid curve shows the DDB Eq.(\ref{ftot}) result for $D_B=0.4, \beta=-0.1$ and $\delta_v=-0.6$.  As we can see, once again we obtain a good agreement with the data. One could ask if the parameters $\delta_v=-0.6$ can be related to any measurement within the simulation. This goes beyond the scope of this paper since it would required an investigation of the mapping between the size of the void and the proto-void. This mapping is well established in the case of spherical evolution ($1.7 R_{Lag}\sim R$). Hence for the void catalogue fitted by our theoretical formula with the spherical barrier prediction $\delta_v=-2.7$, we will test further the consistency of the spherical evolution in section \ref{sdv}. Before we will briefly investigate the effect of the biasing over the identification of voids. 

Finally, unlike the DDB model which is quite flexible on its own, we have found that the ESP model cannot describe the size distribution of voids identified using halos (or sparse DM samples). This is perhaps not so surprising; discrete tracers of voids would likely need to be addressed using a full halo model analysis (e.g. \cite{fp06}). This is beyond the scope of the paper and we leave such an investigation to future work. 

\subsection{The effect of biasing}

 \begin{figure}
\begin{center}
\includegraphics[scale=0.45]{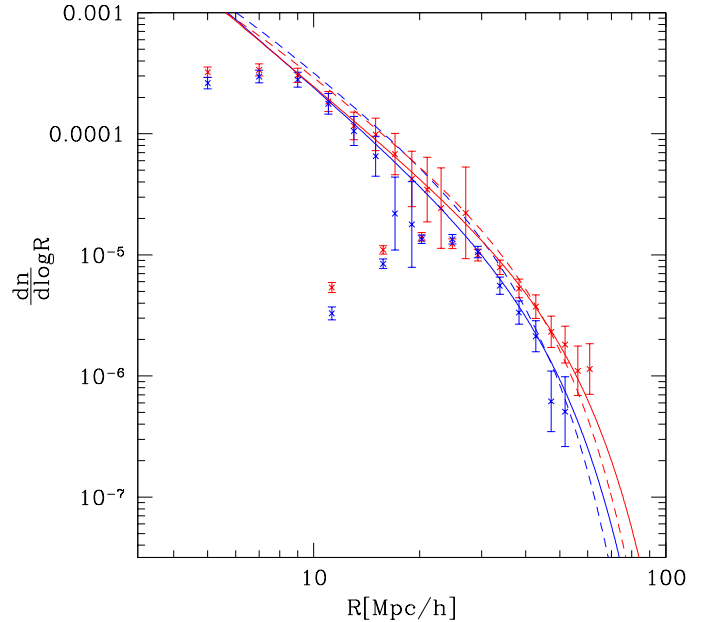}
\caption{Effect of biasing on the void abundance: blue crosses correspond to void identified in the DM density field and red crosses correspond to void identified using halos as biased tracers. The theoretical prediction of SvdW  (dashed curves) and DDB (solid curves) is show for different values of the barrier parameters (see text).}
\label{fig2bis}
\end{center}
\end{figure}

Previously we considered halo positions to study the effect of subsampling matter particles on the abundance of voids. Possibly, biasing could change the void volume function from that of the DM. 
%Physically, we can picture that at a fixed number of tracers (halos or DM particles), the chance to hit twice the same position (modulo a halo radius) is higher in the DM field since we can pick two particles which belong to the same halo. On the contrary, halo positions are separated from each other by at least twice their radius. On the other hand, since not every particle belongs to a halo, some of the randomly selected particles can fill up the empty space between two halos, reducing the abundance of large voids. To see how these two effects compete with each other, 
To see how, we run \zobov\ under the same void selection (core-density $-0.8$) that we previously chose for the halo tracers. We randomly selected DM particles with a total number equal to the number of halos, for each box. In Fig.\ \ref{fig2bis} we can see the resulting void abundance function: red crosses correspond to halos while blue crosses correspond to DM particles for the tracers. As we can see, the effect of biasing is negligible on the abundance of small voids. The abundance of large voids reduces slightly when they are identified by the DM particles. For comparison, the blue dashed curve shows the SvdW prediction for $\delta_v=-0.7$, the blue solid curve is the DDB model with $(D_B=0.4,\delta_v=-0.7,\beta=-0.2)$.\\

To conclude this section, we have shown the difference in the void abundance when the tracer density changes. This allowed us to probe a large range of void radii. We find that without additional post-processing, the $dn/dlogR$ function can be fitted by the SvdW and our theory Eq.(\ref{ftot}) once we rescale $\delta_v=-2.7$ to $\delta_v \sim -0.6$. We note that without further post-processing, a fraction of these voids have an under-density above zero and are strictly speaking not real voids. We find that biased tracers have an effect on large voids. Finally we recover an agreement between the data and theoretical predictions of the void abundance with the usual spherical criteria $\delta_v=-2.7$ for the choice of post-processing case 1, described in the previous section. For the voids thus identified, we next investigate the linear critical underdensities that lead to their formation.

\section{Consistency of the spherical evolution approximation}\label{sdv}
A straightforward way to relate nonspherical voids to the spherical collapse model is to define them around density minima as spheres with underdensity $\Delta_v=-0.8$, as in \cite{jlh13}. Then for all of those voids, we could go back in the initial conditions and test whether the proto-void corresponds to a linear underdensity of  $\delta_v=-2.7$. However, as we already mentioned, including the edge of the void or not makes a significant difference for the void's average density. Therefore, defining voids through a density criterion might be noisy. In this sense, \zobov\ may be more suitable to define voids and compare with observation. Nevertheless, from the theoretical modelling, the spherical shell evolution is generally assumed, and in this section we propose to test whether this assumption is consistent or not. In order to perform this test, we first consider the $z=0$ void volume centroid, a Voronoi-volume-weighted average of particle positions. This center can differ from the one defined using the particle which sits on the minimum density of the void (the `core particle'). Indeed, the minimum of the density profile might not correspond to the minimum of the potential if the surrounding shear field is asymmetric. Therefore for each void, we find the closest particle to its volume centroid. Going back to the initial conditions, we record the underdensity in a sphere centered on this center particle within the Lagrangian radius of the corresponding void.

\begin{figure}
\begin{center}
\includegraphics[scale=0.43]{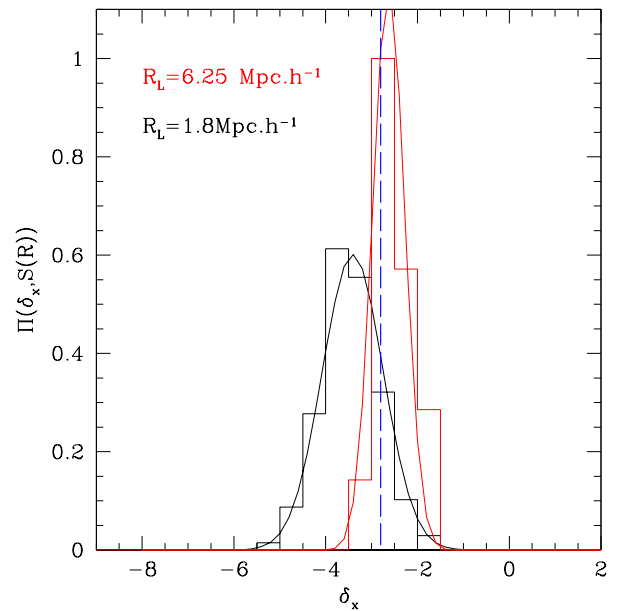}
\caption{Initial overdensities (linearly extrapolated to $z=0$) measured in the $N$-body simulations of regions which produce voids at $z=0$ for two different void radii. Solid lines correspond to Gaussian fits. Blue vertical line shows the spherical model prediction.}\label{fig3}
\end{center}
\end{figure}

In the ideal spherical-shell model, the distribution of this critical underdensity\footnote{the $x$ subscript means that the barrier crosses the smoothed overdensity: $\delta_x\equiv \lbrace  B_v \cap\delta\rbrace$} $\delta\equiv \delta_x$ for all void size  $\Pi(\delta_x,S(R))$ would be a Dirac-delta centred on $\delta_v=-2.7$. Indeed, spherical evolution is fully deterministic. Following the procedure we just described, the PDF we measure is shown in Fig.\ \ref{fig3} for two different void sizes. 

First of all, note that the initial overdensities in Fig.\ \ref{fig3} are negative as expected, which confirms that voids form from underdense regions.  However, there is a significant scatter, indicating that the spherical threshold can not be exactly applied to model the abundance of voids defined as in our previous analysis. Secondly, the mean of the distribution varies for two different voids sizes ($\langle \delta_x \rangle(R=1.8)=-3.4,~\langle\delta_x \rangle (R=6.25)=-2.5)$. While the mean value is close to the spherical prediction for the larger void size, at smaller sizes the implied threshold is significantly more underdense, with a larger scatter. In principle, the scale-dependence of the mean value can be modelled by a negative drift term. Note that this analysis only assumes spherical dynamics on an object-by-object basis; we do not assume any particular statistical model for the void abundance.

In the peaks-based model this would imply a mean value $\avg{\beta_{\rm pk}}\simeq 1$ when using \eqn{esp-stochbar}. More precisely, given the prior $p(\beta_{\rm pk})$ and the peaks calculation of $f(\sigma | \beta_{\rm pk})$, we calculate the volume function $f(\sigma) = \int \der \beta_{\rm pk} p(\beta_{\rm pk}) f(\sigma | \beta_{\rm pk})$ as described in Section 2.2. Then we use Bayes' rule to calculate $p(\beta_{\rm pk} | \sigma) = f(\sigma | \beta_{\rm pk}) p(\beta_{\rm pk}) / f(\sigma)$. This is directly related to $p(\delta_x | \sigma)$ upon using \eqn{esp-stochbar}.

 For the DDB model, the value of $\beta,D_B$ can be inferred using the mapping of Eq.(10) in \cite{ARSC} once we build the PDF of $\delta_x$ at randomly selected positions. Note, however, that different sets of values would not lead to a good agreement with the measured volume function, indicating a possible breakdown of the spherical evolution assumption which we discuss further below.

In addition, we checked that the closest particle to the center void at $z=0$ is at a distance $d$ well below the void radius $R$, to avoid a bias due to a wrong definition of  the void center. Also, we checked that the PDF of $\delta_x$ is almost insensitive to the displacement of this center particle. This is important because if we assume that this particle sits at the minimum of the density, then tracking its displacement tells us if the void forms from an initial underdense peak in Lagrangian space. It also reassures us that we have picked the correct center of the void: if initially this particle is not at the minimum of the potential then it should be pulled out of the center at $z=$0.

Note that alternatively to our protovoid center definition, we could have defined the center in Lagrangian space by averaging over initial positions of particles which belong to the void at $z=0$, weighting each particle in the average by its $z=0$ Voronoi volume. Since most of the uncertainty comes from the void edge, where densities are high; it may be preferable to downweight particles there. However, we expect this definition of the void center to lead to nearly the same results in Fig.\ \ref{fig3}. Indeed, we found that the PDF in Fig.\ \ref{fig3} changes only slightly when the void centers used to measure densities in the initial conditions are displaced to their final positions. This suggests that Fig.\ \ref{fig3} should be insensitive to subtleties in void centering.
%However, we expect this correction to be negligible, since the PDF shown in Fig.\ \ref{fig3} does not change for the center particle which has a small displacement suggesting that the center particle really sits at the minimum of the potential. 
%Note that we expect these displacements to be tiny only for the largest voids; large-scale flows can slightly move small voids and subvoids.

To summarize, this analysis shows that, as expected, voids form from initial underdense regions (see Fig.~\ref{fig3}). Most of the void centers displace over time from their initial location, but the density criterion which leads to their formation is rather uncorrelated to this displacement. Furthermore, the deterministic spherical evolution is apparently not exactly achieved even for larger voids. This could indicate a breakdown of the simple spherical model. It is possible, however, that a modification to the void finder would improve agreement with the model; to test for that, we perform an additional analysis in the next section, selecting spherical-underdensity voids with top-hat average densities $\Delta_v=-0.8$ at $z=0$.

\section{Voids defined as spherical underdensities}
\label{sec:su}
 \begin{figure}
\begin{center}
\includegraphics[scale=0.43]{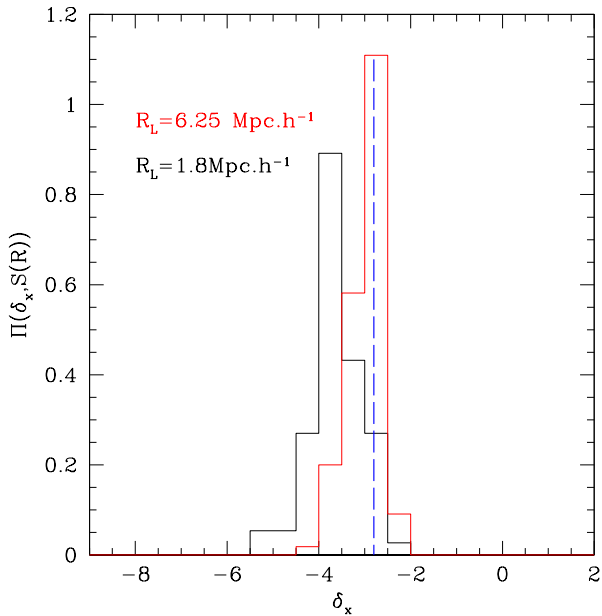}
\caption{Initial overdensities (linearly extrapolated to $z=0$) measured in the $N$-body simulations of regions which produce spherical voids of $\Delta_v=-0.8$ at $z=0$ for two different void radii. Blue vertical line shows the spherical model prediction.}\label{fig3bis}
\end{center}
\end{figure}

Previously, we tested if \zobov\ voids have initial underdensities in agreement with the spherical threshold. Furthermore, the spherical evolution of an isolated underdense patch, leading to a void at $z=0$, gives specific predictions \citep{svdw04}. Assuming: 

a- initial (proto)-voids are spherical TopHats, \\

\noindent this spherical proto-void evolves similarly to an FLRW universe. At shell crossing,  \\

b- initial voids enclose a linear density contrast corresponding to $\delta_v=-2.71$ at z=0;\\

c- voids at z=0 have a density contrast $\Delta_v=-0.8$; and\\

d- voids at z=0 are spherical.\\

This relation between the linear and non-linear underdensity of the spherical shell predicts:\\

e- the radius at z=0 is larger by a factor $q=1.7$ compared to the initial radius ($R_E\sim q R_L$).\\

In this section, we select voids in our catalogues which satisfy points (d) and (c), and test the rest of these predictions. For this purpose, we follow the approach of \cite{jlh13}. We start from \zobov\ void centers (density minima in the Voronoi tessellation) at $z=0$. Then we measure the density at large radius, and move void radius inward until  $\Delta_v=-0.8$. We find that the radius tends to be smaller than the one we find from the \zobov\ output which does not impose a density threshold. Typically the difference is of order $R_E=0.9 R_{E}^{\rm ZOBOV}$. 

Then we repeat the previous analysis: assuming points (a) and (e) are satisfied, we compute the linear underdensities leading to those voids. In this case, we should recover point (b). Fig.\ \ref{fig3bis} shows the result of this new analysis, at two different void sizes. As we can see, the linear underdensity is again not deterministic but it has a distribution which varies as a function of the void size. We report a mean value $\avg{\delta_x(R=1.8)}=-3.6$ and $\avg{\delta_x(R=6.25)}=-2.9$. These are smaller than the previous means ($-3.4$ and $-2.5$, respectively). This decrease is not surprising since on average, the new voids are smaller, comprising mostly the cores of \zobov\ voids. For large voids, the variance is reduced compared to that in Fig.\ \ref{fig3}, indicating that spherical-underdensity voids follow a deterministic evolution better than \zobov\ voids. There is a decreased variance compared to that in Fig.\ \ref{fig3} for the smaller voids as well, although the mean value is still smaller than the spherical model expectation especially for small voids. In the model of \cite{pls12}, the authors find that the Lagrangian patch corresponding to a void would have a
(linearly extrapolated) density contrast below -2.7 – as long as its surrounding has a density to
restrict the expansion (hence forming a wall outside of the void). Since small voids are more
likely to be affected in that model, the average density contrast smoothed over the Lagrangian
scale would be smaller than those of the large voids, matching the trend observed in the  simulation\footnote{We are grateful to the referee for pointing out this explanation}. 

Note that one could probably change point (e) such that $q$ is a function void size, giving $\delta_v=-2.7$ for all voids. Typically, for small voids, this factor would have to be bigger in order to enclose more matter in the initial conditions. However, this would not remove the associated scatter around the mean value. There are a couple of factors that could contribute to this scatter. First, the particle nearest the volume centroid does not necessarily occupy a density minimum in the initial conditions even for an isolated void; it may even be at a small halo. However, such a peak corresponding to a small halo is unlikely to persist in the initial conditions after TopHat-smoothing with the Lagrangian void radius.  Indeed, we did test that using an alternative definition of a void center (the density minimum, instead of volume centroid) produced nearly identical results. Secondly, interactions between voids could lead to a poor correspondence between $z=0$ voids and initial density minima. For example, if two voids that accurately follow spherical evolution are very close to each other, they could merge together, with an undetectably tenuous density ridge between them. Possibly, both \zobov\ and the spherical-underdensity algorithm could report a point near the ridge between them as the void center. This would give a flawed estimate of the radius of the void; also, the reported center could fail to be an underdensity in the initial conditions.\footnote{We thank the referee for pointing out this second effect.}

Despite these possible issues, \citet{jlh13} show that this procedure leads to reasonable predictions of void abundances. Overall it is quite remarkable that the simple spherical model shows an approximate consistency for the large voids, for both \zobov\ and spherical-underdensity voids.

\section{Conclusions}
\label{sec:conclude}

In this paper we develop simple analytical predictions for the void volume function based on the excursion set approach and the peaks formalism which consistently include the effects of filtering with a TopHat in real space. We extended traditional predictions that use a deterministic barrier threshold to a more general class of barriers which are stochastic and sensitive to the size of the voids. The analytical predictions were tested against Monte Carlo simulations of random walks crossing the diffusive drifting barrier and show very good agreement.

We also compared our analytical predictions for the void volume function with numerical results obtained by running the \zobov\ void finder on N-body simulations  with different mass resolution and choice of post-processing. 

\zobov\ is particularly suited to find highly aspherical voids in the cosmic web, as many underdensities are at the present epoch. One (sometimes under-appreciated) aspect of \zobov\ is that the measured void volume function can depend on the choices made on the core threshold parameter. Furthermore many low density contrast voids arise quite frequently in a Poisson process and are likely to be spurious, not corresponding to a physical void. Removing these voids by performing an additional post-processing can change the trend of the void volume function. While this should be kept in mind when analyzing observational results based on \zobov\, this sensitivity comes in only in the high-sampling limit (when discreteness noise produces potentially spurious voids).  

Regardless of post-processing analysis, we find a reasonable agreement between the analytical predictions (SVdW;DDB) and the measured void volume functions once we arbitrary changed the barrier parameters. For one specific choice of post-processing we found that the SVdW, DDB and the peak models reproduced the measured void volume function with the traditional value of $\delta_v=-2.7$. However, the specific parameter values of the stochastic barrier needed to obtain this agreement (at least for the peaks-based model) do not appear to be consistent with our direct measurements of the overdensity threshold in the initial conditions. 
In fact, we showed in a direct measurement in the initial conditions that the usual linear-theory threshold is incompatible with the deterministic value predicted by the spherical model for these \zobov\ voids. These voids have an effective radius less than 10 $\Mpc$ at z=0. This is an important result since current predictions for the abundances of voids defined using compact objects require the linear threshold criterion as an input \citep{fp06}. We also tried a different, spherical-underdensity void finder, as \citet{jlh13} used. With this approach too we find that small voids do not obey spherical evolution in detail; their linearly extrapolated initial densities are generally lower than spherical evolution would predict while larger voids tend to agree with the spherical threshold in average.

%##################
Our work can be extended in several directions: one could compare our prediction with a spherical-underdensity algorithm and see if the corresponding initial threshold tends to the deterministic spherical prediction, introduce a more general barrier as we present here, or perhaps an deeper investigation on the mapping between Lagrangian to Eulerian space could be performed by empirically finding a $q$ factor ($R/q=R_{Lag}$) such that the proto-void underdensity distribution becomes consistent with the linear barrier which enters in the modelling of the void abundance. If such a factor can be found we can already infer that it would have to be smaller than the spherical 1.7 value in order to get a shallower mean underdensity for small voids. Those tests are crucial if we want to have analytical model predictions able to compare to current and future observational void catalogues.

\section*{Acknowledgments}
We thank Yann Rasera for useful discussions about the DEUS N-body simulation, and Ben Wandelt for his reception at IAP. I. Achitouv acknowledges
support from the Trans-Regional Collaborative Research
Center TRR 33 ``The Dark Universe'' of the
Deutsche Forschungsgemeinschaft (DFG). MN is grateful for support from a New Frontiers of Astronomy and Cosmology grant from the Sir John Templeton Foundation.

\bibliography{refVoids}

\begin{thebibliography}{71}
\expandafter\ifx\csname natexlab\endcsname\relax\def\natexlab#1{#1}\fi

\bibitem[{{Achitouv} {et~al}\mbox{.}(2013){Achitouv}, {Rasera}, {Sheth}, \&
  {Corasaniti}}]{ARSC}
{Achitouv} I., {Rasera} Y., {Sheth} R.~K., {Corasaniti} P.~S., 2013, Physical
  Review Letters, 111, 231303

\bibitem[{{Achitouv} {et~al}\mbox{.}(2014){Achitouv}, {Wagner}, {Weller}, \&
  {Rasera}}]{AWWR}
{Achitouv} I., {Wagner} C., {Weller} J., {Rasera} Y., 2014, JCAP, 10, 77

\bibitem[{{Achitouv} \& {Corasaniti}(2012{\natexlab{a}})}]{AC1}
{Achitouv} I.~E., {Corasaniti} P.~S., 2012{\natexlab{a}}, JCAP, 2, 2

\bibitem[{{Achitouv} \& {Corasaniti}(2012{\natexlab{b}})}]{AC2}
{Achitouv} I.~E., {Corasaniti} P.~S., 2012{\natexlab{b}}, \prd, 86, 083011

\bibitem[{{Alimi} {et~al}\mbox{.}(2010){Alimi}, {F{\"u}zfa}, {Boucher},
  {Rasera}, {Courtin}, \& {Corasaniti}}]{Alimi2010}
{Alimi} J.-M., {F{\"u}zfa} A., {Boucher} V., {Rasera} Y., {Courtin} J.,
  {Corasaniti} P.-S., 2010, \mnras, 401, 775

\bibitem[{{Appel} \& {Jones}(1990)}]{aj90}
{Appel} L., {Jones} B.~J.~T., 1990, \mnras, 245, 522

\bibitem[{{Bardeen} {et~al}\mbox{.}(1986){Bardeen}, {Bond}, {Kaiser}, \&
  {Szalay}}]{bbks86}
{Bardeen} J.~M., {Bond} J.~R., {Kaiser} N., {Szalay} A.~S., 1986, \apj, 304, 15
  (BBKS)

\bibitem[{{Bertschinger}(1985)}]{b85}
{Bertschinger} E., 1985, \apjs, 58, 1

\bibitem[{Biswas {et~al}\mbox{.}(2010)Biswas, Alizadeh, \& Wandelt}]{baw10}
Biswas R., Alizadeh E., Wandelt B.~D., 2010, \prd, D82, 023002

\bibitem[{{Blumenthal} {et~al}\mbox{.}(1992){Blumenthal}, {da Costa},
  {Goldwirth}, {Lecar}, \& {Piran}}]{b+92}
{Blumenthal} G.~R., {da Costa} L.~N., {Goldwirth} D.~S., {Lecar} M., {Piran}
  T., 1992, \apj, 388, 234

\bibitem[{{Bond} {et~al}\mbox{.}(1991){Bond}, {Cole}, {Efstathiou}, \&
  {Kaiser}}]{bcek91}
{Bond} J.~R., {Cole} S., {Efstathiou} G., {Kaiser} N., 1991, \apj, 379, 440

\bibitem[{{Chuen Chan} {et~al}\mbox{.}(2014){Chuen Chan}, {Hamaus}, \&
  {Desjacques}}]{Chanetal2014}
{Chuen Chan} K., {Hamaus} N., {Desjacques} V., 2014, ArXiv e-prints

\bibitem[{{Colberg} {et~al}\mbox{.}(2008){Colberg}, {Pearce}, {Foster},
  {Platen}, {Brunino}, {Neyrinck}, {Basilakos}, {Fairall}, {Feldman},
  {Gottl{\"o}ber}, {Hahn}, {Hoyle}, {M{\"u}ller}, {Nelson}, {Plionis},
  {Porciani}, {Shandarin}, {Vogeley}, \& {van de Weygaert}}]{colberg08}
{Colberg} J.~M. {et~al.}, 2008, \mnras, 387, 933

\bibitem[{{Colberg} {et~al}\mbox{.}(2005){Colberg}, {Sheth}, {Diaferio}, {Gao},
  \& {Yoshida}}]{colberg05}
{Colberg} J.~M., {Sheth} R.~K., {Diaferio} A., {Gao} L., {Yoshida} N., 2005,
  \mnras, 360, 216

\bibitem[{{Corasaniti} \& {Achitouv}(2011{\natexlab{a}})}]{CA2}
{Corasaniti} P.~S., {Achitouv} I., 2011{\natexlab{a}}, \prd, 84, 023009

\bibitem[{{Corasaniti} \& {Achitouv}(2011{\natexlab{b}})}]{CA1}
{Corasaniti} P.~S., {Achitouv} I., 2011{\natexlab{b}}, Physical Review Letters,
  106, 241302

\bibitem[{{Courtin} {et~al}\mbox{.}(2011){Courtin}, {Rasera}, {Alimi},
  {Corasaniti}, {Boucher}, \& {F{\"u}zfa}}]{Courtin2011}
{Courtin} J., {Rasera} Y., {Alimi} J.-M., {Corasaniti} P.-S., {Boucher} V.,
  {F{\"u}zfa} A., 2011, \mnras, 410, 1911

\bibitem[{{Croton} {et~al}\mbox{.}(2004){Croton}, {Colless}, {Gazta{\~n}aga},
  {Baugh}, {Norberg}, {Baldry}, {Bland-Hawthorn}, {Bridges}, {Cannon}, {Cole},
  {Collins}, {Couch}, {Dalton}, {de Propris}, {Driver}, {Efstathiou}, {Ellis},
  {Frenk}, {Glazebrook}, {Jackson}, {Lahav}, {Lewis}, {Lumsden}, {Maddox},
  {Madgwick}, {Peacock}, {Peterson}, {Sutherland}, \& {Taylor}}]{Croton04}
{Croton} D.~J. {et~al.}, 2004, \mnras, 352, 828

\bibitem[{{D'Aloisio} \& {Furlanetto}(2007)}]{daloisio2007}
{D'Aloisio} A., {Furlanetto} S.~R., 2007, \mnras, 382, 860

\bibitem[{{D'Amico} {et~al}\mbox{.}(2011){D'Amico}, {Musso}, {Nore{\~n}a}, \&
  {Paranjape}}]{dmnp11}
{D'Amico} G., {Musso} M., {Nore{\~n}a} J., {Paranjape} A., 2011, \prd, 83,
  023521

\bibitem[{{Dubinski} {et~al}\mbox{.}(1993){Dubinski}, {da Costa}, {Goldwirth},
  {Lecar}, \& {Piran}}]{d+93}
{Dubinski} J., {da Costa} L.~N., {Goldwirth} D.~S., {Lecar} M., {Piran} T.,
  1993, \apj, 410, 458

\bibitem[{{Epstein}(1983)}]{e83}
{Epstein} R.~I., 1983, \mnras, 205, 207

\bibitem[{{Fillmore} \& {Goldreich}(1984)}]{fg84}
{Fillmore} J.~A., {Goldreich} P., 1984, \apj, 281, 9

\bibitem[{{Furlanetto} \& {Piran}(2006)}]{fp06}
{Furlanetto} S.~R., {Piran} T., 2006, \mnras, 366, 467

\bibitem[{{Goldberg} {et~al}\mbox{.}(2005){Goldberg}, {Jones}, {Hoyle},
  {Rojas}, {Vogeley}, \& {Blanton}}]{g+05}
{Goldberg} D.~M., {Jones} T.~D., {Hoyle} F., {Rojas} R.~R., {Vogeley} M.~S.,
  {Blanton} M.~R., 2005, \apj, 621, 643

\bibitem[{{Granett} {et~al}\mbox{.}(2008){Granett}, {Neyrinck}, \&
  {Szapudi}}]{GranettEtal2008}
{Granett} B.~R., {Neyrinck} M.~C., {Szapudi} I., 2008, \apjl, 683, L99

\bibitem[{{Hamaus} {et~al}\mbox{.}(2013){Hamaus}, {Wandelt}, {Sutter},
  {Lavaux}, \& {Warren}}]{h+13}
{Hamaus} N., {Wandelt} B.~D., {Sutter} P.~M., {Lavaux} G., {Warren} M.~S.,
  2013, ArXiv e-prints: 1307.2571

\bibitem[{{Hoffman} {et~al}\mbox{.}(1983){Hoffman}, {Salpeter}, \&
  {Wasserman}}]{hsw83}
{Hoffman} G.~L., {Salpeter} E.~E., {Wasserman} I., 1983, \apj, 268, 527

\bibitem[{{Hoyle} {et~al}\mbox{.}(2005){Hoyle}, {Rojas}, {Vogeley}, \&
  {Brinkmann}}]{hrvb05}
{Hoyle} F., {Rojas} R.~R., {Vogeley} M.~S., {Brinkmann} J., 2005, \apj, 620,
  618

\bibitem[{{Hoyle} \& {Vogeley}(2002)}]{hv02}
{Hoyle} F., {Vogeley} M.~S., 2002, \apj, 566, 641

\bibitem[{{Hoyle} \& {Vogeley}(2004)}]{hv04}
{Hoyle} F., {Vogeley} M.~S., 2004, \apj, 607, 751

\bibitem[{{Hunt} \& {Sarkar}(2010)}]{hs10}
{Hunt} P., {Sarkar} S., 2010, \mnras, 401, 547

\bibitem[{{Jennings} {et~al}\mbox{.}(2013){Jennings}, {Li}, \& {Hu}}]{jlh13}
{Jennings} E., {Li} Y., {Hu} W., 2013, ArXiv e-prints: 1304.6087

\bibitem[{{Kamionkowski} {et~al}\mbox{.}(2009){Kamionkowski}, {Verde}, \&
  {Jimenez}}]{kvj09}
{Kamionkowski} M., {Verde} L., {Jimenez} R., 2009, JCAP, 1, 10

\bibitem[{{Kauffmann} \& {Fairall}(1991)}]{kf91}
{Kauffmann} G., {Fairall} A.~P., 1991, \mnras, 248, 313

\bibitem[{{Kirshner} {et~al}\mbox{.}(1981){Kirshner}, {Oemler}, {Schechter}, \&
  {Shectman}}]{koss81}
{Kirshner} R.~P., {Oemler}, Jr. A., {Schechter} P.~L., {Shectman} S.~A., 1981,
  \apjl, 248, L57

\bibitem[{{Lacey} \& {Cole}(1993)}]{LaceyCole}
{Lacey} C., {Cole} S., 1993, \mnras, 262, 627

\bibitem[{{Lam} \& {Sheth}(2009)}]{LamSheth2009}
{Lam} T.~Y., {Sheth} R.~K., 2009, \mnras, 398, 2143

\bibitem[{{Lam} {et~al}\mbox{.}(2009){Lam}, {Sheth}, \& {Desjacques}}]{lsd09}
{Lam} T.~Y., {Sheth} R.~K., {Desjacques} V., 2009, \mnras, 399, 1482

\bibitem[{{Lavaux} \& {Wandelt}(2010)}]{lw10}
{Lavaux} G., {Wandelt} B.~D., 2010, \mnras, 403, 1392

\bibitem[{{Lavaux} \& {Wandelt}(2012)}]{LavauxWandelt2012}
{Lavaux} G., {Wandelt} B.~D., 2012, \apj, 754, 109

\bibitem[{{Maggiore} \& {Riotto}(2010{\natexlab{a}})}]{MR1}
{Maggiore} M., {Riotto} A., 2010{\natexlab{a}}, \apj, 711, 907

\bibitem[{{Maggiore} \& {Riotto}(2010{\natexlab{b}})}]{MR2}
{Maggiore} M., {Riotto} A., 2010{\natexlab{b}}, \apj, 717, 515

\bibitem[{{Melchior} {et~al}\mbox{.}(2013){Melchior}, {Sutter}, {Sheldon},
  {Krause}, \& {Wandelt}}]{Melchior+13}
{Melchior} P., {Sutter} P.~M., {Sheldon} E.~S., {Krause} E., {Wandelt} B.~D.,
  2013, ArXiv e-prints: 1309.2045

\bibitem[{{Musso} \& {Sheth}(2012)}]{ms12}
{Musso} M., {Sheth} R.~K., 2012, \mnras, 423, L102

\bibitem[{{Neyrinck}(2008)}]{Neyrinck2008}
{Neyrinck} M.~C., 2008, \mnras, 386, 2101

\bibitem[{{Neyrinck} {et~al}\mbox{.}(2005){Neyrinck}, {Gnedin}, \&
  {Hamilton}}]{NeyrinckEtal2005}
{Neyrinck} M.~C., {Gnedin} N.~Y., {Hamilton} A.~J.~S., 2005, \mnras, 356, 1222

\bibitem[{{Pan} {et~al}\mbox{.}(2012){Pan}, {Vogeley}, {Hoyle}, {Choi}, \&
  {Park}}]{Pan12}
{Pan} D.~C., {Vogeley} M.~S., {Hoyle} F., {Choi} Y.-Y., {Park} C., 2012,
  \mnras, 421, 926

\bibitem[{{Paranjape} {et~al}\mbox{.}(2012){Paranjape}, {Lam}, \&
  {Sheth}}]{pls12}
{Paranjape} A., {Lam} T.~Y., {Sheth} R.~K., 2012, \mnras, 420, 1648

\bibitem[{{Paranjape} \& {Sheth}(2012)}]{ps12}
{Paranjape} A., {Sheth} R.~K., 2012, \mnras, 426, 2789

\bibitem[{{Paranjape} {et~al}\mbox{.}(2013){Paranjape}, {Sheth}, \&
  {Desjacques}}]{psd13}
{Paranjape} A., {Sheth} R.~K., {Desjacques} V., 2013, \mnras, 431, 1503

\bibitem[{{Park} \& {Lee}(2007)}]{pl07}
{Park} D., {Lee} J., 2007, \prl, 98, 081301

\bibitem[{{Patiri} {et~al}\mbox{.}(2006{\natexlab{a}}){Patiri},
  {Betancort-Rijo}, \& {Prada}}]{pbp06}
{Patiri} S.~G., {Betancort-Rijo} J., {Prada} F., 2006{\natexlab{a}}, \mnras,
  368, 1132

\bibitem[{{Patiri} {et~al}\mbox{.}(2006{\natexlab{b}}){Patiri},
  {Betancort-Rijo}, {Prada}, {Klypin}, \& {Gottl{\"o}ber}}]{p+06}
{Patiri} S.~G., {Betancort-Rijo} J.~E., {Prada} F., {Klypin} A.,
  {Gottl{\"o}ber} S., 2006{\natexlab{b}}, \mnras, 369, 335

\bibitem[{{Peacock} \& {Smith}(2000)}]{ps00}
{Peacock} J.~A., {Smith} R.~E., 2000, \mnras, 318, 1144

\bibitem[{{Pisani} {et~al}\mbox{.}(2013){Pisani}, {Lavaux}, {Sutter}, \&
  {Wandelt}}]{Pisani13}
{Pisani} A., {Lavaux} G., {Sutter} P.~M., {Wandelt} B.~D., 2013, ArXiv
  e-prints:1306.3052

\bibitem[{{Platen} {et~al}\mbox{.}(2007){Platen}, {van de Weygaert}, \&
  {Jones}}]{PlatenEtal2007}
{Platen} E., {van de Weygaert} R., {Jones} B.~J.~T., 2007, \mnras, 380, 551

\bibitem[{{Press} \& {Schechter}(1974)}]{ps74}
{Press} W.~H., {Schechter} P., 1974, \apj, 187, 425

\bibitem[{{Rasera} {et~al}\mbox{.}(2010){Rasera}, {Alimi}, {Courtin}, {Roy},
  {Corasaniti}, {F{\"u}zfa}, \& {Boucher}}]{yann}
{Rasera} Y., {Alimi} J.-M., {Courtin} J., {Roy} F., {Corasaniti} P.-S.,
  {F{\"u}zfa} A., {Boucher} V., 2010, in American Institute of Physics
  Conference Series, Vol. 1241, American Institute of Physics Conference
  Series, {Alimi} J.-M., {Fu{\"o}zfa} A., eds., pp. 1134--1139

\bibitem[{{Robertson} {et~al}\mbox{.}(2009){Robertson}, {Kravtsov}, {Tinker},
  \& {Zentner}}]{Roberton}
{Robertson} B.~E., {Kravtsov} A.~V., {Tinker} J., {Zentner} A.~R., 2009, \apj,
  696, 636

\bibitem[{{Ryden}(1995)}]{Ryden1995}
{Ryden} B.~S., 1995, \apj, 452, 25

\bibitem[{{Sahni} {et~al}\mbox{.}(1994){Sahni}, {Sathyaprakah}, \&
  {Shandarin}}]{sss94}
{Sahni} V., {Sathyaprakah} B.~S., {Shandarin} S.~F., 1994, \apj, 431, 20

\bibitem[{{Schaap} \& {van de Weygaert}(2000)}]{SchaapVandeWeygaert2000}
{Schaap} W.~E., {van de Weygaert} R., 2000, \aap, 363, L29

\bibitem[{{Seljak}(2000)}]{Seljak2000}
{Seljak} U., 2000, \mnras, 318, 203

\bibitem[{{Sheth} \& {van de Weygaert}(2004)}]{svdw04}
{Sheth} R.~K., {van de Weygaert} R., 2004, \mnras, 350, 517 (SvdW)

\bibitem[{{Sutter} {et~al}\mbox{.}(2012){Sutter}, {Lavaux}, {Wandelt}, \&
  {Weinberg}}]{Sutter12}
{Sutter} P.~M., {Lavaux} G., {Wandelt} B.~D., {Weinberg} D.~H., 2012, \apj,
  761, 44

\bibitem[{{Teyssier}(2002)}]{Ramses}
{Teyssier} R., 2002, \aap, 385, 337

\bibitem[{{Tinker} \& {Conroy}(2009)}]{tc09}
{Tinker} J.~L., {Conroy} C., 2009, \apj, 691, 633

\bibitem[{{van de Weygaert} \& {Platen}(2011)}]{vdwp11}
{van de Weygaert} R., {Platen} E., 2011, International Journal of Modern
  Physics Conference Series, 1, 41

\bibitem[{{van de Weygaert} \& {van Kampen}(1993)}]{vdwvk93}
{van de Weygaert} R., {van Kampen} E., 1993, \mnras, 263, 481

\bibitem[{{Zhang} \& {Hui}(2006)}]{ZhangHui2006}
{Zhang} J., {Hui} L., 2006, \apj, 641, 641

\end{thebibliography}

\label{lastpage}

\end{document}